%% file: main.tex
\documentclass[runningheads]{llncs}
\usepackage[T1]{fontenc}
\usepackage{graphicx}
\usepackage{booktabs}
\usepackage{multirow}
\usepackage{tabularx}
\usepackage{amsmath}
\begin{document}
\graphicspath{ {./images/} }

\title{Agent-Based Exploration of Recommendation Systems in Misinformation Propagation}



\author{
Lise Jakobsen\inst{1}\orcidID{0000-0000-0000-0001} \and
Anna Johanne Holden Jacobsen\inst{1}\orcidID{0000-0000-0000-0002} \and
\"{O}nder G\"{u}rcan\inst{2}\orcidID{0000-0001-6982-5658} \and
\"{O}zlem \"{O}zg\"{o}bek\inst{1}\orcidID{0000-0000-0000-0002}
}

\authorrunning{L. Jakobsen, A. H. Jacobsen, \"{O}. G\"{u}rcan and \"{O}. \"{O}zg\"{o}bek}

\institute{
    Norwegian University of Science and Technology (NTNU), Trondheim, Norway \\
    \email{\{lise.jakobsen, anna.jacobsen, ozlem.ozgobek\}@ntnu.no}
    \and
    NORCE Norwegian Research Centre AS, Kristiansand, Norway\\
    \email{ongu@norceresearch.no}
}

\maketitle              
\begin{abstract}
This study uses agent-based modeling to examine the impact of various recommendation algorithms on the propagation of misinformation on online social networks. We simulate a synthetic environment consisting of heterogeneous agents, including regular users, bots, and influencers, interacting through a social network with recommendation systems. We evaluate four recommendation strategies: popularity-based, collaborative filtering, and content-based filtering, along with a random baseline. Our results show that popularity-driven algorithms significantly amplify misinformation, while item-based collaborative filtering and content-based approaches are more effective in limiting exposure to fake content. Item-based collaborative filtering was found to perform better than previously reported in related literature. These findings highlight the role of algorithm design in shaping online information exposure and show that agent-based modeling can be used to gain realistic insight into how misinformation spreads. 

\end{abstract}
\section{Introduction}
\label{Intro}

\input{content/Introduction}

\section{Background and Related work}
\label{BackAndRelated}
\input{content/RelatedWork}

\section{Methodology}
\label{Methods}

\input{content/Methodology}


\section{Results}
\label{Results}

\input{content/Results}

\section{Discussion}
\label{Discussion}
\input{content/Discussion}

\section{Conclusion and Future Work}
\label{Conclusion}

\input{content/Conclusion}

\bibliographystyle{ieeetr}
\bibliography{bibliography/bibtext}

\end{document}

%% file: content/Introduction.tex
The significant presence of social media today has made it a primary source of information, which has revolutionized the way individuals consume and interact with content and news \cite{HermidaShareLikeRecommend}. However, this digital transformation has also facilitated the rapid spread of misinformation, a phenomenon with severe implications for public discourse, democracy, and social trust \cite{Masip2018300}. Misinformation can alter the balance of authenticity of the news ecosystem, exposing people to equal amounts of true and fake news, making it more difficult for consumers to differentiate which news is trustworthy and which is not \cite{ShuDataMining2017}.

Misinformation on social media is often propagated by malicious users and bots, both of which play significant roles in amplifying false information \cite{Ruan2020,Shao2018}. 
Social media platforms are unable to control what users post until these actors have already broken the community guidelines. 
Misinformation is further shown to be reinforced through social media's use of recommendation systems \cite{PathakUnderstanding2023}.  
Even though the platforms can not control the malicious user behavior, they can control how the recommendation algorithms (RAs) curate the user news feeds and therefore have a great influence on what users are exposed to \cite{gausen2022abm}.

Recommendation systems are designed to enhance user experience by personalizing content selection to user preferences. While useful for dealing with content churn and enhancing user experience in social media and news domains, this personalization can inadvertently promote the formation of filter bubbles and echo chambers, where users are repeatedly exposed to similar viewpoints \cite{Lunardi2020}. Currently, there is limited focus within recommendation systems on identifying, restricting, or correcting misinformation content. The consequences of failing to address the role of RAs in misinformation propagation are prominent \cite{PathakUnderstanding2023}. Beyond the immediate societal harms, such as increased polarization and erosion of trust in institutions, the unchecked spread of misinformation undermines the credibility of digital platforms themselves. As these systems continue to influence public opinion on critical issues, from elections to public health, there is an ethical incentive for researchers to develop solutions that help mitigate the propagation of misinformation in recommendation systems.

Looking into how the different RAs affect the flow of misinformation online is therefore important to evaluate new approaches for mitigation. Previous research focusing on the effect of RAs on the spread of misinformation provides a good baseline \cite{Fernandez2024,PathakUnderstanding2023}. Our approach differentiates itself with the use of agent-based modeling (ABM), in which we simulate a social network with realistic features and investigate how recommendation systems affect the spread of misinformation on social networks. 

Based on this observation, in this study, we focus on the extent to which different RAs affect misinformation spread. We also investigate how agent-based modeling can help us gain realistic insight into how misinformation spreads in a social network that employs RAs. 
The contributions of our study are as follows:

\begin{itemize}
    \item Developed a detailed agent-based model\footnote{The source code will be made publicly available if the paper is accepted.} that simulates a social network with heterogeneous agents, including regular users, bots, influencers. The model is implemented using Python, the Mesa framework, and the LensKit library for recommender systems - —building on best‑practice guidelines for re‑implementable ABMs \cite{Gurcan2023}.
    \item Compared how various recommendation strategies— popularity-based, collaborative filtering, content-based, and random—affect misinformation spread.
    \item Integrated temporal dynamics and feedback loops in the simulation to realistically model evolving misinformation spread and algorithmic influence over time.
    \item Applied evaluation metrics like infection rate, Misinformation Ratio Difference (MRD), and Misinformation Count (MC) to quantify the effects of different algorithms on misinformation dissemination.
    \item  Found that item-based collaborative filtering performed better than expected, limiting misinformation spread, contrary to previous studies.
\end{itemize}

 



The rest of the paper is organized as follows: Section \ref{BackAndRelated} reviews background and related work; Section \ref{Methods} describes the simulation framework and evaluation metrics; Section \ref{Results} presents and analyzes the results; Section \ref{Discussion} discusses implications, prior work, and limitations; and Section \ref{Conclusion} concludes with future research directions.

%% file: content/RelatedWork.tex
\subsection{Agent-Based Modeling (ABM)}

ABM offers a framework for analyzing the complex interactions between users, algorithms, and information flow on social media platforms \cite{gausen2022abm,Gurcan2013}. Unlike static models, ABM simulates autonomous agents with behaviors that are governed by simple rules, and where interactions give an understanding of complex system-level dynamics.

Many agent-based models are based on epidemiological diffusion frameworks, such as the SIR (Susceptible-Infected-Recovered) model, which effectively captures the viral nature of information spread in networks \cite{Govindankutty2024}. Several studies simulate misinformation spread using epidemiological diffusion models such as SIR due to its ability to mimic viral behavior in networks. These models serve as the foundation for more complex agent-based approaches used in recent research \cite{Soni2024} These foundational models are extended in ABM by embedding cognitive, social, and algorithmic components into each agent's behavior. Agents can be influenced by their preferences, social neighbors, previous experiences, or exposure to content, making ABM particularly suitable for simulating phenomena such as misinformation cascades, opinion shifts, and the emergence of echo chambers.

Recent studies have used ABM to investigate how misinformation spreads and how recommendation systems could contribute to or mitigate such spread \cite{gausen2022abm,Li2024,Ahmed2024}. In our work, we build on these insights to model not only individual user behavior but also the dynamic feedback loops introduced by different recommendation strategies and evolving network structures.

\subsection{Recommendation Algorithms and Misinformation}

Recommendation systems are technologies designed to suggest relevant content to users based on their preferences, behaviors, or interactions \cite{Ko2022}.
The risk of recommendation systems amplifying the spread of misinformation has been highlighted in several studies. Pathak et al. \cite{PathakUnderstanding2023} demonstrated how RAs, particularly those based on popularity and networks, are particularly likely to recommend misleading content. One key concern is the feedback loop between user behavior and the algorithm. If users engage with misinformation, the RAs interpret it as positive feedback and then continue to promote similar content. Over time, this dynamic leads to the emergence of filter bubbles and echo chambers, where users are increasingly exposed to information that confirms their existing beliefs. This was further explored by Lunardi et al. \cite{Lunardi2020}, who proposed a metric to quantify echo chambers and demonstrated how certain collaborative filtering algorithms can propagate misinformation among like-minded users. 

\cite{Fernandez2024} provided guidelines for adapting existing algorithms, highlighting the need to go beyond accuracy and incorporate fairness, transparency, and misinformation awareness into recommendation logic. Our work further builds upon the work of Fernandez and Pathak, but with a focus on simulating the RAs impact on misinformation spread using ABM.

\subsection{Social Networks and Misinformation}
Social media has become the main news source for our society \cite{ShuDataMining2017}. Although they offer the opportunity for open communication, they also facilitate the rapid viral spread of misinformation due to their architectural design and social dynamics.

The 2016 U.S. presidential election and the COVID-19 pandemic highlighted how misinformation can influence public opinion and behavior. During this period, more than one million tweets were related to fake news, and the most shared fake election stories generated more than 8.5 million interactions on Facebook, including shares, reactions, and comments \cite{ShuDataMining2017,Zhou2020}. 

Social networks amplify misinformation spread through features such as algorithmic timelines, retweets, and likes. When social media platforms optimize for user engagement, users are at risk of only being exposed to content that aligns with their previous beliefs, as that is the content they most likely interact with. The tendency that a person prefers to accept information that confirms their preexisting beliefs is referred to as confirmation bias \cite{ShuDataMining2017}. This further leads to filter bubbles and echo chambers, isolating users from opposite viewpoints and resulting in increased polarization.

%% file: content/Methodology.tex
This section describes the simulation design used to investigate the spread of misinformation in social networks. ABM incorporates various types of agents, social network topologies, RAs, and content dynamics. The model has been implemented in Python using the Mesa framework, and the RAs are built using LensKit where fitting \cite{MesaKazil,LenskitEkstrand}, or written by us. 

\subsection{Simulation Framework}
The agent population consists of 90\% regular users, 7\% bots, and 3\% influencers. Agents vary in their activity, naivety, influence, and probability of spreading misinformation. All agent characteristics are displayed in Table \ref{tab:agent_attributes}.

\begin{table}[h!]
\centering
\caption{Agent Attributes by Type}
\label{tab:agent_attributes}
\renewcommand{\arraystretch}{1} 
\begin{tabularx}{\textwidth}{|>{\raggedright\arraybackslash}X|>{\centering\arraybackslash}X|>{\centering\arraybackslash}X|>{\centering\arraybackslash}X|}
\hline
\textbf{Attribute} & \textbf{Regular User} & \textbf{Bot Agent} & \textbf{Influencer Agent} \\
\hline
Activity Probability & 0.1--0.7 (mean=0.3) & 0.4--0.9 (mean=0.7) & 0.3--0.8 (mean=0.5) \\
\hline
Naivety & Moderate & High & Moderate \\
\hline
Activity Pattern & 1--3 peak times/day & High activity & Strategic posting times \\
\hline
Post Misinformation Probability & Regular & High & Low \\
\hline
State & \multicolumn{3}{c|}{Susceptible, Exposed, Infected (S, E, I)} \\
\hline
Content Preferences & \multicolumn{3}{c|}{Personalized preference vector} \\
\hline
Feed & \multicolumn{3}{c|}{Content from followed agents and recommendations} \\
\hline
Recommendations & \multicolumn{3}{c|}{Top-$N$ items from RA} \\
\hline
\end{tabularx}
\end{table}

We model a synthetic social network with a directed graph using NetworkX. Agent connectivity is based on the cosine similarity between individual preference vectors, creating a network topology that reflects users having a stronger likelihood of following each other if they have similar preferences. Influencers have on average about 20 times more followers than regular users, while bots have fewer followers but exhibit persistent activity patterns designed to maximize exposure.

Agents are tracked using a SEI (Susceptible, Exposed, Infected) diffusion model, which builds upon the SIR epidemic model \cite{bissettEpidemic,hengSEIR}. Our modified version tracks misinformation interaction progression. Agents begin as \textbf{Susceptible (S)}, transition to \textbf{Exposed (E)} when encountering fake content, and become \textbf{Infected (I)} once they engage with it. After 40 time steps, they may return to the states S or E, depending on whether or not their feeds include misinformation.

Content in the simulation is initialized either by the model in the beginning, or by users posting new content. Engagement decays over time using an exponential decay function, influencing the content's likelihood of being interacted with. This implementation of time dynamics was inspired by \cite{Lotito2021}. Whether an agent shares content is determined by the content’s similarity to their preferences, their naivety, and the item's current engagement score. Bots are highly prone to sharing misinformation, while influencers are more selective. Content characteristics are detailed in Table \ref{tab:misinformation_model}.

\begin{table}[htbp]
\centering
\caption{Misinformation Model Components}
\label{tab:misinformation_model}
\small
\begin{tabular}{|l|p{0.75\columnwidth}|}
\hline
\textbf{Component} & \textbf{Description} \\
\hline
Content ID & Unique identifier for each content item \\
\hline
\texttt{isFake} flag & Boolean value indicating if the content is misinformation \\
\hline
Topic Vector & Vector representation of the content's thematic topic \\
\hline
Initial Engagement & Real news: 1.0; Fake news: 1.5 \\
\hline
Engagement Decay & Exponential decay rate: 0.1 \\
\hline
Content Evaluation & Cosine similarity with agent preference vector $\times$ agent credibility $\times$ content engagement (capped at 1.5x) \\
\hline
Creation step & Set to model.steps when initialized \\
\hline
\end{tabular}
\normalsize
\end{table}

Content diffusion occurs through both network shares and RA-driven recommendations. When shared, content propagates to followers; when recommended, it is pushed into user feeds based on the RA type. The interplay between agent traits, content dynamics, and network topology drives the emerging spread of misinformation.

\subsection{Recommendation Algorithms}
We explore a variety of the classical recommendation algorithms. For our experiment, we evaluated how the different recommendation strategies affect the misinformation spread and the formation of echo chambers. 


\begin{itemize}
    \item \textbf{Non-personalized recommendation}: Includes popularity-based recommendations based on recent engagement and virality and random recommendations for cold-start scenarios or as a baseline.
    \item \textbf{Collaborative filtering recommendation}: Suggests items based on user similarity (user-based) or item similarity (item-based) using collaborative filtering algorithms implemented with the LensKit Library \cite{LenskitEkstrand}.
    \item \textbf{Content-based filtering recommendation}: Ranks items using cosine similarity between content topic vectors and the agent's preference vector.
\end{itemize}

\subsection{Metrics}
In our investigation, we use metrics that will let us capture the performance of RAs and the impact of the misinformation. No metric alone can address all the important criteria of a recommendation system comprehensively \cite{AdaptiveWebBook}. Therefore, to measure our specific goals, we will be looking at the following metrics: misinformation spread (infection rate), which is how many agents are infected over time. This will give us insight into what recommendation algorithm is most prone to spreading misinformation and how fast it happens. This is calculated by detecting the number of agents in the state "Infected" for each timestep, as can be seen in Equation \ref{eq:msp}. 

\begin{equation} \label{eq:msp}
\text{MSP} = \left( \frac{N_{\text{infected}}}{N_{\text{total}}} \right) \times 100\%
\end{equation}

Where \(N_{\text{infected}}\) is the number of infected agents, and \(N_{\text{total}}\)
is the total number of agents.
To further evaluate the RAs performance, we calculate the Misinformation Ratio Difference (MRD), which is the difference between the misinformation rate in recommendations and the total misinformation \cite{Fernandez2024}. 

A positive value will indicate that the RA is over-recommending misinformation with a risk of amplifying echo chambers. A negative value will mean that the RA is under-recommending misinformation, which can indicate a possible bias control. We use Misinformation Count (MC) to measure the count of misinformation items recommended to each user on average \cite{Fernandez2024}. A higher MC value will indicate that there is more misinformation included in the recommendations. 

\subsection{Experimental protocol} \label{sec:experimental_protocol}
To evaluate the influence of recommendation algorithms on misinformation spread and echo chamber formation, we performed a series of agent-based simulations under consistent initial conditions. Each simulation was initialized with the same fixed network of regular users, bots, and influencers and a balanced pool of content that includes a controlled percentage of fake news.

Deciding on the values of the parameters became a decision based on assumptions, due to the lack of scientific research published on the ratio between users and average followers and the exact percentage of misinformation online. Consequently, we made all decisions based on our own perception of the ratios of users, influencers, bots, followers, and misinformation on social media platforms. We chose to generate 10 recommendations per user agent in each timestep to ensure sufficiently large content feeds, thus increasing the likelihood of user interaction. This allowed us to observe changes in the network earlier and reduce the total number of timesteps needed in the simulation. The chosen parameters can be seen in Table \ref{tab:parameters}.

\begin{table}[h]
\centering
\caption{Parameters used in the experiments.}
\begin{tabular}{ll}
\textbf{Parameter} & \textbf{Value} \\
\hline
Timesteps & 600 \\
Number of Users & 200 \\
Average Followers & 6 \\
Initial News Amount & 400 \\
Misinformation Percentage & 10\% \\
Bot Percentage & 7\% \\
Influencer Percentage & 3\% \\
Recommendations per Step & 10 \\
\end{tabular}

\label{tab:parameters}
\end{table}

To run the experiment, we utilized the Mesa BatchRunner API to run parallel executions of the experiment \cite{MesaKazil}. Each run consisted of running the simulation using all the different RAs but with the same conditions. The experiment runs for five iterations due to the implicit randomness of the model to ensure valid results. 
After the experiment is finished, we run a plotting script that uses the stored CSV file of data collected, and then outputs summary plots of the metrics previously described for further analysis.

%% file: content/Results.tex


\subsubsection{The Effect of Recommendation Algorithms on Misinformation Spread:}

\begin{figure}
    \centering
    \includegraphics[width=1\linewidth]{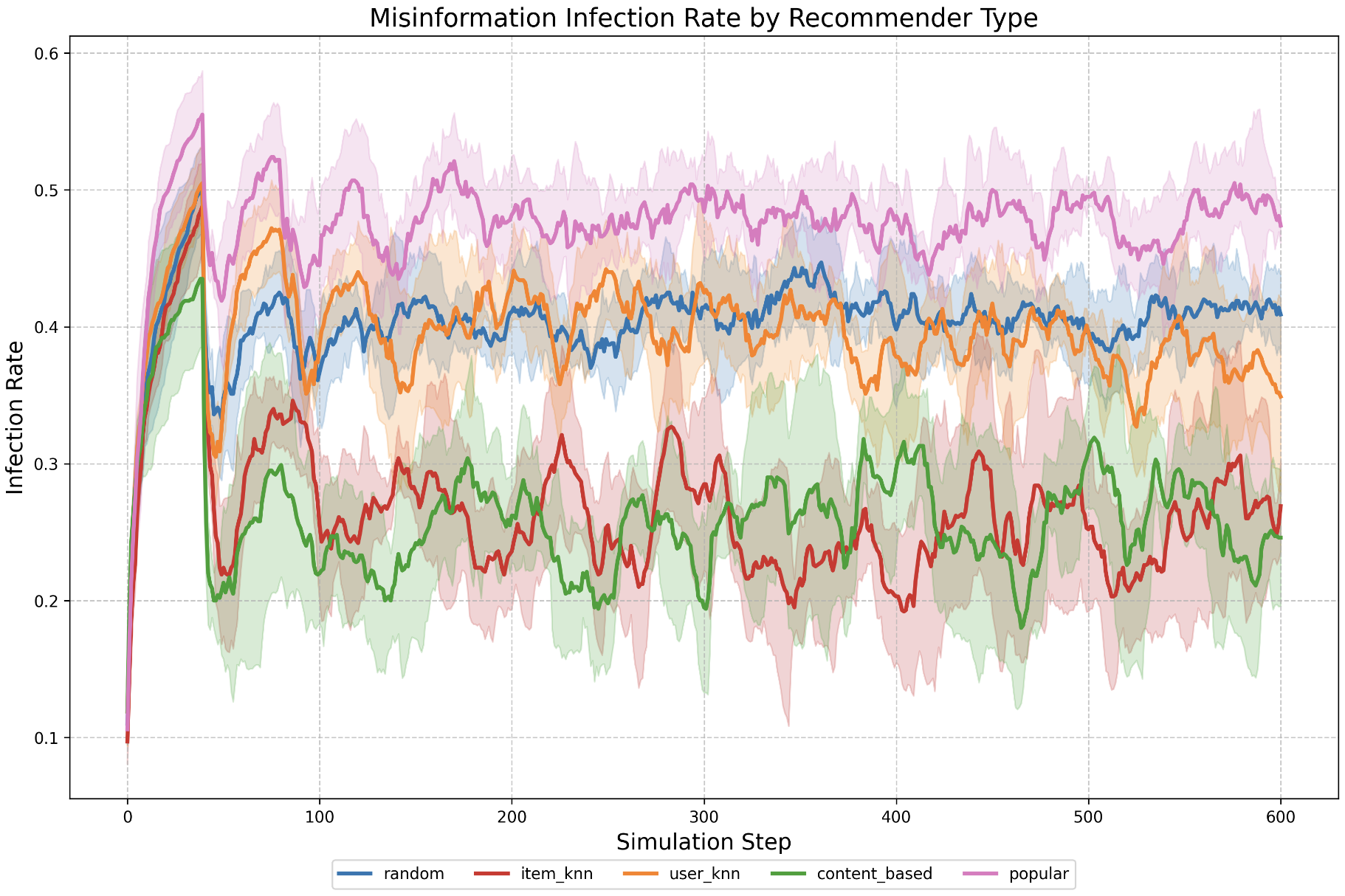}
    \label{fig:infection_comparison}
    \caption{Infection rate over time}
\end{figure}

\begin{figure}
    \centering
    \includegraphics[width=1\linewidth]{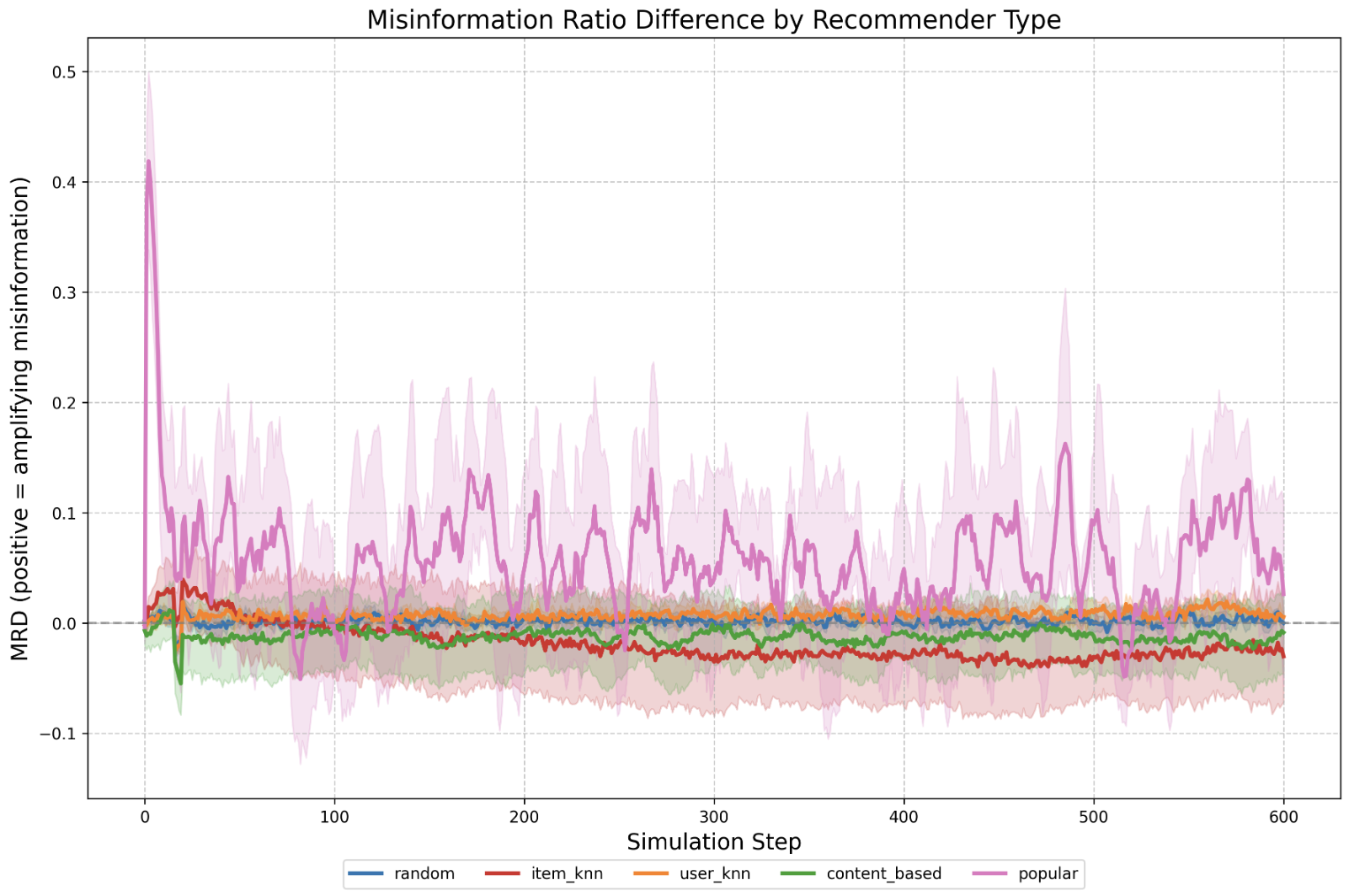}
    \caption{MRD over time}
    \label{fig:mrd_comparison}
\end{figure}

\begin{figure}
    \centering
    \includegraphics[width=1\linewidth]{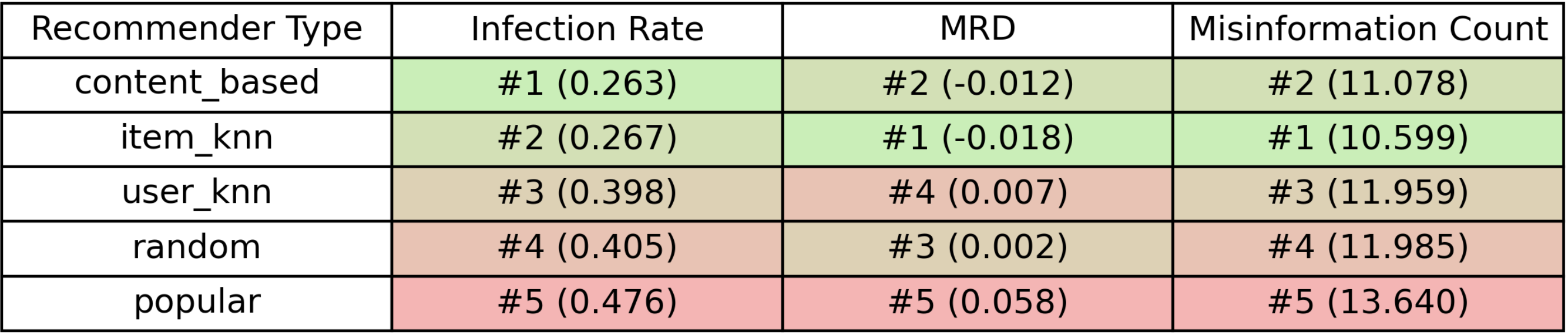}
    \caption{Summary of metrics on average}
    \label{fig:ranking_table}
\end{figure}

We present the overall results on to what extent different RAs affect misinformation spread. Figure \ref{fig:ranking_table} compares the overall average metrics regarding each RA, where each one is ranked from best (\#1 - least propagating of misinformation) to worst (\#5 - most propagating of misinformation). Figure \ref{fig:infection_comparison} shows the percentage of agents that have been infected with misinformation in each simulation step, and Figure \ref{fig:mrd_comparison} shows how each algorithm amplifies misinformation in recommendations on average during each simulation step.

Results show that the popularity-based model exhibited the highest sustained infection rate, as well as recommending the most fake news. The popularity-based algorithm also frequently shows positive MRD spikes, as seen in Figure \ref{fig:mrd_comparison}, revealing that such algorithms tend to over-recommend misinformation during these spikes. This behavior is consistent with the algorithm's non-personalized design, which recommends content that receives high levels of engagement to most users. In other words, if misinformation content gains popularity, the popularity-based algorithm is likely to further amplify its reach. Misinformation is also more engaging than regular news content, as we have implemented in our model \cite{Munusamy2024}, further enhancing its easy access to virality.

The random algorithm, our baseline, was ranked second worst overall in the three metrics, with the user-knn algorithm as a close third. Looking further at the actual scores in Figure \ref{fig:ranking_table}, user-knn had a higher MRD than random and only performed better than random in MC with a score of 0.026, and achieved just a 0.07\% improvement on the infection rate. This indicates that user-knn performs almost the same as the baseline algorithm. User-knn performs significantly worse than the other collaborative filtering method, item-knn. When users who are more susceptible to misinformation due to higher naivety levels are grouped together by user-knn, the algorithm will recommend content that these similar users have engaged with. Thus, the algorithm can create clusters of naive users where misinformation can propagate more easily. 

Item-based collaborative filtering and the content-based algorithm consistently minimized both exposure and infection rates, with a negative MRD, suggesting a filtering effect against misinformation. Both performed quite similarly regarding infection rates, and content based only performed slightly worse regarding MC. Both of these algorithms focus on the content itself when making recommendations, which might explain why they perform so similarly. In Figure \ref{fig:mrd_comparison} a small decreasing slope can be noticed in the item-knn algorithm. This may be explained by how, over time, more interactions are added to the user-item matrix, and the item-knn algorithm will be able to build more accurate item similarity matrices, thereby becoming better at distinguishing high- and low-quality content. This creates a positive feedback loop that gradually shifts recommendations away from misinformation.

\subsubsection{Insights From Agent-Based Modeling on Misinformation Spread in a Social Network that Employs RAs:}

Our results show that ABM can give us valuable insights regarding how recommendation systems affect misinformation in online social networks. 
Our simulation model is able to capture complex behaviors by modeling agents with varying characteristics, such as naivety, activity levels, influence, and content preferences. 
Using a simulation that runs over time gives us information on how the algorithm's performance changes having a reinforcement loop, which is interesting to observe in Figure \ref{fig:infection_comparison} and \ref{fig:mrd_comparison}. 
Looking at Figure \ref{fig:infection_comparison}  and Figure \ref{fig:mrd_comparison}, the spikes of high MRD values from the popular algorithm represent a quite realistic occurrence, where misinformation becomes very popular in certain periods, but they do not go viral for very long. 

Our simulation is as realistic as the complexity we have managed to integrate, and the realism is also limited by the synthetic essence of the model's foundation. By getting results we find plausible and explainable, together with observing phenomenons that capture real-life occurrences, we can at least conclude that using agent-based modeling can give us some realistic insight into how misinformation spreads in a social network.


%% file: content/Discussion.tex

Our agent-based simulation provides information on how different types of recommendation systems influence the propagation of misinformation on online social networks. By incorporating feedback loops, our model more closely mirrors real-world conditions, where user interactions with content influence subsequent recommendations. This dynamic aspect enhances the realism of our simulation and allows us to observe how algorithmic biases may evolve over time. Unlike the primarily static evaluation frameworks used in the studies by Pathak and Fernandez, our inclusion of feedback loops allows us to model self-reinforcing behavior patterns, such as the dominance of misinformation in a user’s feed. This aligns with concerns raised in the literature on engagement-optimized platforms \cite{Fernandez2024,PathakUnderstanding2023}.

In line with previous findings \cite{PathakUnderstanding2023,Fernandez2024}, which demonstrated that popularity-based algorithms most aggressively propagate misinformation, our results further substantiate this statement. This was a result we expected beforehand, due to the algorithm's nature and because misinformation is more engaging than regular news. The popular algorithm also leads to less diversity across all content recommended to users, and many users will receive the same popular content, meaning misinformation can reach a large amount of the social network in a short amount of time. As our results align with other researchers, we conclude that in a time-sensitive analysis, the non-personalized popular algorithm still has the worst and highest impact on misinformation spread in a social network.  

However, our findings diverge from prior work in one important aspect. In Pathak's study, collaborative filtering methods - both user-based and item-based - tended to recommend a higher volume of misinformation \cite{PathakUnderstanding2023}. We found that item-based collaborative filtering is not only resistant to misinformation propagation, but also consistently outperforms other algorithms in minimizing both exposure and infection. Rather than being susceptible to co-engagement bias, the item-based model in our simulation effectively limits misinformation through its focus on item similarity, suggesting it may help filter out low-quality or misleading content clusters.

Interestingly, our study and Pathak agree on the strength of content-based algorithms in limiting misinformation \cite{PathakUnderstanding2023}. In both analyses, content-based filtering yields lower misinformation metrics, likely due to its reliance on semantic matching rather than user behavior patterns or viral trends.

Running experiments that include time dynamics can be shown to align well with other research performed on static datasets. The most significant difference in our simulation study remains how the item-based collaborative filtering method helped restrict the spread of misinformation. 

\paragraph{Limitations}
While our simulation captures key findings of misinformation spread on social media, there are also certain limitations. First, agents in the simulation follow fixed rules with predefined personality types. This simplification removes the behavioral diversity and adaptive learning that are observed in real social media users. This limits the model's ability to reflect nuanced opinion dynamics. Additionally, user agents are characterized only by user preferences, without consideration of demographic or sociocultural factors such as age, gender, marital status, or occupation. The abstraction may cause our model to miss significant links between how these attributes affect how users behave online. 

All user preferences, topic vectors, and network structures are generated synthetically and remain static throughout the simulation. As a result, the model does not capture the temporal evolution in user interests or the emergence of topic trends, such as breaking news. Content creation is tightly linked with user preferences, ignoring the variability and creativity of real human-generated content. Engagement with content is based solely on topic similarity thresholds, overlooking other important factors such as emotional appeal, presentation quality, or source credibility.

The social network itself is also static throughout the simulation, missing important dynamics such as network growth, unfollowing behavior, or shifting community structures. Platform-specific features such as trending topics and algorithmic feed boosts are also not modeled, which could significantly influence misinformation diffusion.

Finally, while our implementation of recommendation algorithms reflects real-world systems, they are simplified and manually created abstractions. In reality, social media platforms incorporate hybrid recommendation systems that combine elements of content-based, collaborative, and popularity-based approaches. Evaluating these algorithms in isolation may therefore not fully capture the complexity in the actual social network environments.

%% file: content/Conclusion.tex
The results from our experiments tell us that the non-personalized popular algorithm has the worst implications for misinformation propagation in our social network. This finding aligns with previous studies. 
Another result substantiating other research is how content-based algorithms perform quite well regarding the spread of misinformation, likely due to its focus on content similarity and preference dynamics, and its lack of linking naive users together in clusters. 
In our study, the performance of the item-knn algorithm is the result that differs the most from other evaluated studies. This method was the most effective in limiting the spread of misinformation. These results indicate that using an agent-based simulation approach is a good alternative for gaining realistic insight into how recommendation systems affect misinformation propagation. Additionally, it provides valuable information on how algorithms influence the spread over time. 

This study opens several promising opportunities for future research. Enhancing behavioral realism through adaptive, memory-equipped agents with evolving beliefs could better reflect user behavior over time. Also, integrating real-world datasets would help validate the model and improve parameter grounding. Further on, supporting dynamic preferences would more accurately simulate long-term engagement patterns. Future work could also explore richer network dynamics, including temporal evolution, platform-specific rules, and advanced recommendation algorithms such as hybrid systems with controllable diversity. 
Recent work on LLM‑enhanced ABMs \cite{Gurcan2024} outlines concrete steps in this direction.

Our ABM highlights the potential of simulation-based studies to uncover the mechanisms by which RAs shape the information landscape. Although limited by abstractions, the model provides a flexible approach to test intervention strategies and explore algorithmic trade-offs. Future enhancements could bring us closer to practical tools for RA design that balance personalization, diversity, and misinformation resilience.

%% file: main.bbl
\begin{thebibliography}{10}

\bibitem{HermidaShareLikeRecommend}
A.~Hermida, F.~Fletcher, D.~Korell, and D.~Logan, ``Share, like, recommend:
  Decoding the social media news consumer,'' {\em Journalism Studies}, vol.~13,
  no.~5-6, p.~815 – 824, 2012.
\newblock Cited by: 462.

\bibitem{Masip2018300}
P.~Masip, J.~Suau-Martínez, and C.~Ruiz-Caballero, ``Questioning the selective
  exposure to news: Understanding the impact of social networks on political
  news consumption,'' {\em American Behavioral Scientist}, vol.~62, no.~3,
  p.~300 – 319, 2018.
\newblock Cited by: 36.

\bibitem{ShuDataMining2017}
K.~Shu, A.~Sliva, S.~Wang, J.~Tang, and H.~Liu, ``Fake news detection on social
  media: A data mining perspective,'' {\em ACM SIGKDD Explor. Newsl.}, vol.~19,
  08 2017.

\bibitem{Ruan2020}
Z.~Ruan, B.~Yu, X.~Shu, Q.~Zhang, and Q.~Xuan, ``The impact of malicious nodes
  on the spreading of false information,'' {\em Chaos}, vol.~30, no.~8, 2020.
\newblock Cited by: 16.

\bibitem{Shao2018}
C.~Shao, G.~L. Ciampaglia, O.~Varol, K.-C. Yang, A.~Flammini, and F.~Menczer,
  ``The spread of low-credibility content by social bots,'' {\em Nature
  Communications}, vol.~9, no.~1, 2018.
\newblock Cited by: 776; All Open Access, Gold Open Access, Green Open Access.

\bibitem{PathakUnderstanding2023}
R.~Pathak, F.~Spezzano, and M.~S. Pera, ``Understanding the contribution of
  recommendation algorithms on misinformation recommendation and misinformation
  dissemination on social networks,'' {\em ACM Trans. Web}, vol.~17, Oct. 2023.

\bibitem{gausen2022abm}
A.~Gausen, W.~Luk, and C.~Guo, ``Using agent-based modelling to evaluate the
  impact of algorithmic curation on social media,'' {\em J. Data Inf. Qual},
  vol.~15, 2022.

\bibitem{Lunardi2020}
G.~M. Lunardi, G.~M. Machado, V.~Maran, and J.~P.~M. de~Oliveira, ``A metric
  for filter bubble measurement in recommender algorithms considering the news
  domain,'' {\em Applied Soft Computing}, vol.~97, p.~106771, 2020.

\bibitem{Fernandez2024}
M.~Fernandez, A.~Bellogín, and I.~Cantador, ``Analysing the effect of
  recommendation algorithms on the spread of misinformation,'' in {\em
  Proceedings of the 16th ACM Web Science Conference}, WebSci '24, (New York,
  NY, USA), pp.~159--169, Association for Computing Machinery, 2024.

\bibitem{Gurcan2023}
\"{O}nder G\"{u}rcan, T.~Szczepanska, and P.~Antosz, ``A guide to
  re‑implementing agent‑based models: Experiences from the humat model,''
  in {\em Advances in Social Simulation, ESSA 2023}, Springer, 2023.

\bibitem{Gurcan2013}
\"{O}nder G\"{u}rcan, O.~Dikenelli, and C.~Bernon, ``A generic testing
  framework for agent-based simulation models,'' {\em Journal of Simulation},
  vol.~7, no.~3, pp.~183--201, 2013.

\bibitem{Govindankutty2024}
S.~Govindankutty and S.~Gopalan, ``Epidemic modeling for misinformation spread
  in digital networks through a social intelligence approach,'' {\em Scientific
  Reports}, vol.~14, p.~19100, 2024.

\bibitem{Soni2024}
H.~Kumar~Soni, S.~Sharma, and G.~Sinha, eds., {\em Text and Social Media
  Analytics for Fake News and Hate Speech Detection}.
\newblock Chapman and Hall/CRC, 1st~ed., 2024.

\bibitem{Li2024}
X.~Li, Y.~Xu, Y.~Zhang, and E.~C. Malthouse, ``Large language model-driven
  multi-agent simulation for news diffusion under different network
  structures,'' 2024.
\newblock Preprint.

\bibitem{Ahmed2024}
G.~Ahmed, ``Analysis and simulation of misinformation spread in social
  networks: A hybrid stochastic-deterministic approach with neab model,'' {\em
  Communications on Applied Nonlinear Analysis}, vol.~32, pp.~550--560, 09
  2024.

\bibitem{Ko2022}
H.~Ko, S.~Lee, Y.~Park, and A.~Choi, ``A survey of recommendation systems:
  Recommendation models, techniques, and application fields,'' {\em
  Electronics}, vol.~11, 01 2022.

\bibitem{Zhou2020}
X.~Zhou and R.~Zafarani, ``A survey of fake news: Fundamental theories,
  detection methods, and opportunities,'' {\em ACM Computing Surveys}, vol.~53,
  09 2020.

\bibitem{MesaKazil}
J.~Kazil, D.~Masad, and A.~Crooks, ``Utilizing python for agent-based modeling:
  The mesa framework,'' in {\em Social, Cultural, and Behavioral Modeling}
  (R.~Thomson, H.~Bisgin, C.~Dancy, A.~Hyder, and M.~Hussain, eds.), (Cham),
  pp.~308--317, Springer International Publishing, 2020.

\bibitem{LenskitEkstrand}
M.~D. Ekstrand, ``Lenskit for python: Next-generation software for recommender
  systems experiments,'' in {\em Proceedings of the 29th ACM International
  Conference on Information \& Knowledge Management}, CIKM '20, (New York, NY,
  USA), p.~2999–3006, Association for Computing Machinery, 2020.

\bibitem{bissettEpidemic}
K.~Bissett, J.~Cadena, M.~Khan, {\em et~al.}, ``Agent-based computational
  epidemiological modeling,'' {\em Journal of the Indian Institute of Science},
  vol.~101, pp.~303--327, 2021.

\bibitem{hengSEIR}
K.~Heng and C.~L. Althaus, ``The approximately universal shapes of epidemic
  curves in the susceptible-exposed-infectious-recovered (seir) model,'' {\em
  Scientific Reports}, vol.~10, p.~19365, 2020.

\bibitem{Lotito2021}
Q.~F. Lotito, D.~Zanella, and P.~Casari, ``Realistic aspects of simulation
  models for fake news epidemics over social networks,'' {\em Future Internet},
  vol.~13, no.~3, 2021.

\bibitem{AdaptiveWebBook}
P.~Brusilovsky, A.~Kobsa, and W.~Nejdl, {\em The Adaptive Web - Methods and
  Strategies of Web Personalization}.
\newblock Springer Berlin/Heidelberg, 01 2007.

\bibitem{Munusamy2024}
S.~Munusamy, K.~Syasyila, A.~A.~H. Shaari, M.~A. Pitchan, M.~R. Kamaluddin, and
  R.~Jatnika, ``Psychological factors contributing to the creation and
  dissemination of fake news among social media users: a systematic review,''
  {\em BMC Psychology}, vol.~12, no.~1, 2024.
\newblock Open Access.

\bibitem{Gurcan2024}
\"{O}nder G\"{u}rcan, ``Llm‑augmented agent‑based modelling for social
  simulations: Challenges and opportunities,'' in {\em Proc.\ 3rd Intl.\
  Conf.\ on Hybrid Human‑AI (HHAI 2024)}, 2024.

\end{thebibliography}
